\documentclass[aps,twocolumn,superscriptaddress]{revtex4-2}
\usepackage{amsmath,amssymb,bm}
\usepackage{graphicx}
\usepackage{xcolor}

\begin{document}


\title{Magnetic-field-induced corner states in quantum spin Hall insulators} 

\author{Sergey S.~Krishtopenko}
\affiliation{Laboratoire Charles Coulomb (L2C), UMR 5221 CNRS-Universit\'{e} de Montpellier, F- 34095 Montpellier, France}

\author{Fr\'{e}d\'{e}ric Teppe}
\email[]{frederic.teppe@umontpellier.fr}
\affiliation{Laboratoire Charles Coulomb (L2C), UMR 5221 CNRS-Universit\'{e} de Montpellier, F- 34095 Montpellier, France}
\date{\today}

\begin{abstract}
We address the problem of magnetic-field-induced corner states in quantum spin Hall insulators (QSHIs) beyond the particle-hole-symmetric limit. Starting from a realistic low-energy model for zinc-blende semiconductor quantum wells (QWs), we derive the effective edge Hamiltonian in the form of a Dirac Hamiltonian with two magnetic-field-dependent mass terms, whose structure depends on the crystallographic orientation of the edge and of the magnetic-field orientation. Our \emph{analytical} results show that magnetic-field-induced corner states are most naturally understood as in-gap bound states of the effective edge theory, controlled by the relative configuration of the edge mass vectors rather than, in general, as higher-order topological corner modes protected by a stable bulk invariant. We demonstrate that, although mirror-graded winding numbers can be defined and quantized for certain crystallographic configurations, the existence of magnetic-field-induced corner states is not restricted to regimes in which these bulk invariants are well defined. Finally, we argue that even without higher-order topological protection these corner states may remain spectrally robust under weak perturbations as isolated in-gap quasiparticle excitations.
\end{abstract}

\maketitle

\section{\label{Sec:Int} Introduction}
Higher-order topological insulators (HOTIs) represent an important extension of the modern classification of topological phases~\cite{cst1,cst2,cst3,cst4,cst5}.
While conventional (first-order) $m$-dimensional topological insulators host gapless boundary states at their $(m-1)$-dimensional boundaries~\cite{cst6,cst49,cst50,cst8,cst9,cst10}, an $n$th-order phase may exhibit boundary signatures at co-dimension $n$, such as corner or hinge modes. To date, higher-order topological phases have been observed in bismuth~\cite{cst11}, WTe$_2$~\cite{cst12}, bismuth-halide chains~\cite{cst13}, and Bi$_4$Br$_4$~\cite{cst14}, and predicted in many other systems including bulk crystals and quantum wells~\cite{cst15,cst17,cst18,cst19,cst16,cst20}. Related concepts have also been realized in photonic~\cite{cst26,cst27,cst28} and acoustic crystals~\cite{cst29,cst30,cst31}, topolectrical circuits~\cite{cst32,cst33,cst34,cst34b}, and superconductors~\cite{cst21,cst22,cst23,cst24,cst25}.

At the same time, in-gap corner states may also emerge in systems that are not described as stable HOTIs within the standard symmetry-based classification.
A prominent example is provided by quantum spin Hall insulators (QSHIs), where breaking time-reversal symmetry by a magnetic field gaps the helical edge modes and may generate localized states at the junction of two gapped edges~\cite{cst47a00,cst47,cst48}.  Many theoretical studies of magnetic-field-induced corner states rely on lattice models with an additional energy-reflection symmetry, such as particle-hole or chiral symmetry, which pins the corner-state energy to the middle of the bulk gap and strongly constrains symmetry-allowed local perturbations~\cite{cst47a00,cst47,cst48}. Realistic QSHIs, however, typically do not possess such spectral symmetries. In particular, QSHIs based on zinc-blende semiconductor quantum wells (QWs), including HgTe/CdTe~\cite{cst49,cst50} and InAs/Ga(In)Sb QW structures~\cite{cst51,cst52,cst53,cst53b,cst53c,cst53c2,cst53d,cst53e}, exhibit pronounced electron-hole asymmetry and, importantly, may contain bulk inversion-asymmetry (BIA) and interface inversion-asymmetry (IIA) contributions, which modify the Zeeman coupling of the edge states~\cite{cst54,cst55,cst56,cst56b}.

In this work, we address the general problem of magnetic-field-induced corner-localized bound states in QSHIs. Our analysis starts from a realistic continuum description of zinc-blende semiconductor QWs and the corresponding low-energy bulk Hamiltonian. From this model, we derive the effective Hamiltonian for the helical edge states, taking into account BIA and IIA as well as the Zeeman coupling to an external magnetic field. The resulting edge theory has the form of a Dirac Hamiltonian with two magnetic-field-dependent mass terms, whose structure depends on the crystallographic orientation of the edge. In this sense, the corner states obtained below are represented as a realization of the generalized Jackiw-Rebbi mechanism~\cite{cst42}, emerging at the junction of two edges with different magnetic-field-induced mass terms of the effective edge Dirac Hamiltonian.

A central outcome of the present analysis is that, in realistic zinc-blende QSHIs, magnetic-field-induced corner states are most naturally understood as bound states of the effective edge theory rather than, in general, as higher-order topological corner modes protected by a stable bulk invariant. Their existence is controlled by the relative configuration of the magnetic-field-induced edge masses and not, in general, by mirror-symmetry-protected bulk topology~\cite{cst47a00,cst47,cst48}. For completeness, we also evaluate the mirror-graded winding numbers~\cite{bookSM1q,cst60}, which are known to characterize higher-order topological phases in the presence of particle-hole or chiral symmetry. We demonstrate that although mirror-graded winding numbers can be defined and take quantized values for certain crystallographic orientations of the edges and magnetic field, the existence of corner states in zinc-blende QSHIs is not restricted to parameter regimes in which these bulk invariants are well defined. In particular, the corner states persist over a much broader range of edge orientations and magnetic-field directions, including situations in which mirror-graded winding numbers vanish or cannot be defined.

The resulting picture also clarifies the role of perturbations. Although the corner states discussed in this work are not protected by a stable bulk higher-order invariant, they may nevertheless remain spectrally robust under weak perturbations. In this case, the relevant notion of robustness is not topological protection in the strict sense of bulk-boundary correspondence, but the persistence of the corner-state quasiparticle excitation inside the edge gap with parameters renormalized by the perturbation. This viewpoint naturally connects the problem of corner-state spectral robustness to the quasiparticle description developed for disorder- and interaction-induced topological phase transitions in QSHI systems~\cite{TAI3,TAI4}. While a systematic analysis of perturbation effects on magnetic-field-induced corner states is beyond the scope of the present work, the analytical approach demonstrated here provides a natural starting point for such studies.

\section{Bulk and edge Hamiltonian for zinc-blende QSHIs}
\begin{figure}
\includegraphics [width=1.0\columnwidth, keepaspectratio] {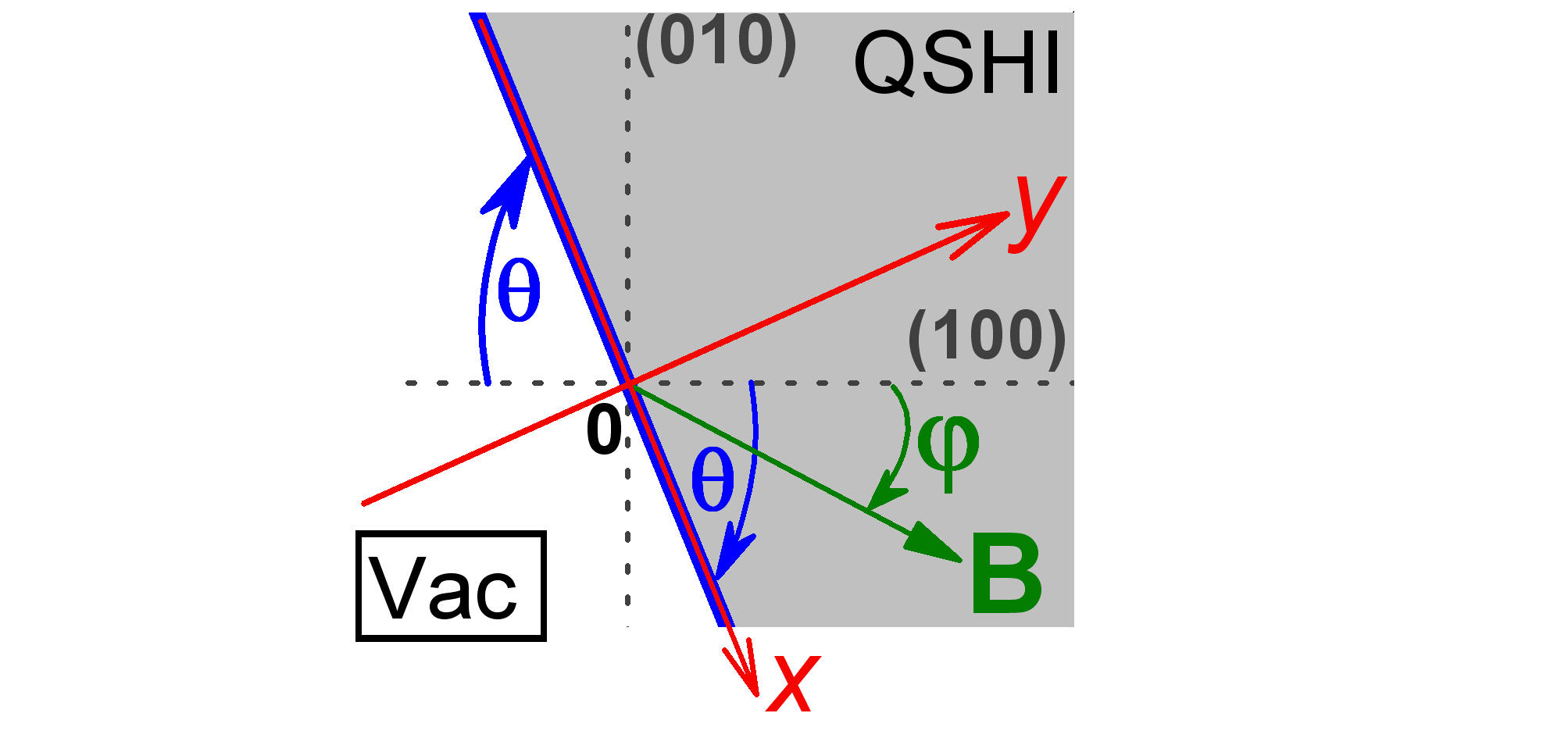} 
\caption{\label{Fig:1} The layout of the coordinate axes (in red) and orientation the magnetic field $B$ (in green) with respect to the main crystallographic axes shown by the vertical and horizontal dashed lines. The QW edge (in blue) considered in the text is oriented along $x$ axis. The positive values of $\theta$ and $\varphi$ correspond to a clockwise rotation around $z$ axis.}
\end{figure}

First, we focus on prototype zinc-blende semiconductor QW with symmetric heteropotential grown along the $z||[001]$ axis. The low-energy Hamiltonian of such systems for the electron-like $E1$ and heavy-hole-like $H1$ subbands, written in the basis $|E1{\uparrow}\rangle$, $|H1{\uparrow}\rangle$, $|E1{\downarrow}\rangle$, $|H1{\downarrow}\rangle$, has the form~\cite{cst49}:
\begin{equation}
\label{eq:1}
H_{\mathrm{2D}}(\mathbf{k})=\begin{pmatrix}
H_{\mathrm{BHZ}}(\mathbf{k}) & -i\Delta{e^{-2i\theta}}\sigma_y \\
i\Delta{e^{2i\theta}}\sigma_y & H_{\mathrm{BHZ}}^{*}(-\mathbf{k})\end{pmatrix},
\end{equation}
where asterisk stands for complex conjugation, $\mathbf{k}=(k_x,k_y)$ is the momentum in the plane, and $H_{\mathrm{BHZ}}(\mathbf{k})=\epsilon_{k}+d_a(\mathbf{k})\sigma_a$. Here, $\sigma_a$ are the Pauli matrices acting in the ``basis'' space, $\epsilon_{k}=\mathbb{C}-\mathbb{D}k^2$, $d_1(\mathbf{k})=\mathbb{A}k_x$, $d_2(\mathbf{k})=\mathbb{A}k_y$, $d_3(k)=\mathbb{M}-\mathbb{B}k^2$ and $k^2=k_x^2+k_y^2$. The mass parameter $\mathbb{M}$ describes the inversion between the electron-like \emph{E}1 and the hole-like \emph{H}1 subbands: $\mathbb{M}>0$ corresponds to a trivial state, while $\mathbb{M}<0$ is for QSHI~\cite{cst49}. The non-diagonal terms in $H_{\mathrm{2D}}(\mathbf{k})$ proportional to $\Delta$ describe the joint effect of bulk inversion asymmetry (BIA) of the unit cell of zinc-blende semiconductors~\cite{cst54} and the interface inversion asymmetry (IIA) of the QW~\cite{cst55,cst56}. Experimental results known from the literature show that the values of $\Delta$ are small in HgTe/CdTe QWs~\cite{A1,A2,A3,A4,A5,A6,A6b,A7}, while they can be relatively large in InAs/GaInSb-based heterostructures~\cite{A8,A9,A10,A11,A11b,A12,A13}. In $H_{\mathrm{2D}}(\mathbf{k})$, $\theta$ is the angle between the $x$ axis and the (001) crystallographic direction (see Fig.~\ref{Fig:1}). This coordinate system allows one to consider the edge states with different crystallographic orientations of the edge. To obtain $H_{\mathrm{2D}}(\mathbf{k})$ in Eq.~(\ref{eq:1}) from a Hamiltonian with $\theta=0$, together with a clock-wise rotation of the coordinate system along $z$ axis, it is also necessary to perform a unitary transformation of the Hamiltonian given by $U=\exp(-i\theta{J_\xi})$. Here, $J_\xi$ is a diagonal matrix with the elements $(1/2,3/2,-1/2,-3/2)$ corresponding to the momentum of the basis states $|E1{\uparrow}\rangle$, $|H1{\uparrow}\rangle$, $|E1{\downarrow}\rangle$, $|H1{\downarrow}\rangle$, respectively~\cite{cst49}.

In order to take into account effect of in-plane magnetic field $\mathbf{B}$, one has to make the Peierls substitution $\mathbf{k}\rightarrow\mathbf{k}-(e/c\hbar)\mathbf{A}$ in $H_{\mathrm{2D}}(\mathbf{k})$ and to add the Zeeman Hamiltonian:
\begin{equation}
\label{eq:3}
H_{\mathrm{Z}}=\dfrac{\mu_B}{2}\begin{pmatrix}
0 & 0 & g_e^{||}B_{+} & 0\\
0 & 0 & 0 & g_h^{||}e^{-i4\theta}B_{-}\\
g_e^{||}B_{-} & 0 & 0 & 0\\
0 & g_h^{||}e^{i4\theta}B_{+} & 0 & 0 \end{pmatrix},
\end{equation}
where $B_{\pm}=B_{x}{\pm}iB_{y}$, $\mu_B$ is the Bohr magneton, $g_e^{||}$ and $g_h^{||}$ are the in-plane g-factors of the $E1$ and $H1$ subbands, resulting from the bare electron g-factor and the interaction with the remote electron and hole subbands~\cite{Wbook}. Since, one can always choose the vector potential gauge so that $\mathbf{A}||z$ for in-plane magnetic field, we can always restrict ourselves to considering only the Zeeman term omitting the Peierls substitution for $k_x$ and $k_y$. The angular dependence of $H_{\mathrm{Z}}$ on $\theta$ in the chosen coordinate system is obtained in the same way as for $H_{\mathrm{2D}}(\mathbf{k})$. As clear, $\Delta$ and $g_h^{||}$ are the terms breaking rotational symmetry of $H_{\mathrm{2D}}(\mathbf{k})$ and $H_{\mathrm{Z}}$. They both origin from the contribution of the $\Gamma_8$ bulk band of zinc-blende semiconductors~\cite{Wbook}. In the following, the parameters $\Delta$ and $g_h^{||}$ allow us to control the strength of crystal-symmetry-induced anisotropies. Setting $\Delta=0$ and $g_h^{||}=0$ restores an effectively rotationally symmetric model, which provides a useful reference limit for identifying the role of crystal symmetry in the formation
of corner states in the prototype (001) QW.

To analyze the existence of corner states, we first need to obtain a 1D edge Hamiltonian. To derive an effective 1D low-energy edge Hamiltonian, we consider a semi-infinite plane $y>0$ with open-boundary conditions at $y=0$ as shown in Fig.~\ref{Fig:1}. Then, we solve the eigenvalue problem for $H_{\mathrm{2D}}(\mathbf{k})$ at $\mathbb{M}<0$ to find the edge wave-functions at $k_x=0$. Finally, to derive the edge Hamiltonian, one should
successively project the part of $H_{\mathrm{2D}}(\mathbf{k})$ with non-zero $k_x$ and $H_{\mathrm{Z}}$ onto the calculated edge wave-functions. The procedure described above is fairly standard, so we do not present its details here. Up to our knowledge, in the presence of non-zero $\Delta$, it was first implemented by Durnev~\emph{et~al.}~\cite{cst56} After routine mathematics, the part of the edge Hamiltonian independent of $\mathbf{B}$ at small values of $k_x$ is written as
\begin{equation}
\label{eq:2}
H_{\mathrm{edge}}^{(0)}(k_x)=\varepsilon_0+\hbar{v_F}k_x{s}_z+\mathcal{O}\left(k_x^2\right),
\end{equation}
where $s_a$ ($a=x,y,z$) corresponds to the Pauli matrices acting on spin degree of freedom,
$\varepsilon_0=\mathbb{C}-\eta\mathbb{M}$ and $\hbar{v_F}$ is defined as~\cite{cst56b}
\begin{equation}
\label{eq:2b}
\hbar{v_F}=\mathbb{A}\sqrt{1-\eta^2}\Omega,
\end{equation}
where $\eta=\mathbb{D}/\mathbb{B}$ and
\begin{equation}
\label{eq:2c}
\Omega=\dfrac{|\mathbb{M}|\sqrt{1-\eta^2}}{\sqrt{\left(1-\eta^2\right)\mathbb{M}^2+\Delta^2}}.
\end{equation}
Note that the non-linear terms in $k_x$ arise in Eq.~(\ref{eq:2}) due to the presence of $\Delta$. They are all vanishing if $\Delta=0$~\cite{cst56b}.

Similarly, the Zeeman part of the edge Hamiltonian can be presented in the form~\cite{cst56}
\begin{equation}
\label{eq:4}
H_{\mathrm{edge}}^{(\mathbf{B})}=\dfrac{\mu_B}{2}\sum_{a,b=x,y}g_{ab}s_{a}B_{b},
\end{equation}
where the edge g-factor tensor $g_{ab}$ depends on the edge orientation with the following non-zero components:
\begin{eqnarray}
\label{eq:5}
g_{xx}=g_1\cos^{2}2\theta+g_2\sin^{2}2\theta,~\nonumber\\
g_{yy}=g_1\sin^{2}2\theta+g_2\cos^{2}2\theta,~\nonumber\\
g_{xy}=g_{yx}=\dfrac{1}{2}\left(g_1-g_2\right)\sin{4\theta}.
\end{eqnarray}
Here, $g_1$ and $g_2$ are two constants written as~\cite{cst56b}
\begin{eqnarray}
\label{eq:6}
g_{1}=g_e^{||}\dfrac{1-\eta}{2}+g_h^{||}\dfrac{1+\eta}{2},~~~~\nonumber\\
g_{2}=\left(g_e^{||}\dfrac{1-\eta}{2}-g_h^{||}\dfrac{1+\eta}{2}\right)\Omega.
\end{eqnarray}
As clear, if one neglects the terms resulting from the crystal symmetry of zinc-blende semiconductors, $g_1=g_2$, and the g-factor of the edges states becomes independent of the edge orientation.

Assuming that the magnetic field $\mathbf{B}$ is oriented at an angle $\varphi$ to the [100] crystallographic direction as shown in Fig.~\ref{Fig:1}, Eq.~(\ref{eq:4}) is rewritten as
\begin{equation}
\label{eq:7}
H_{\mathrm{edge}}^{(\mathbf{B})}=M_x{s}_{x}+M_y{s}_{y},
\end{equation}
where
\begin{eqnarray}
\label{eq:8}
M_x=\dfrac{E_{Z}}{2}\left[g_{+}\cos(\theta-\varphi)+g_{-}\cos(3\theta+\varphi)\right],~\nonumber\\
M_y=\dfrac{E_{Z}}{2}\left[g_{+}\sin(\theta-\varphi)+g_{-}\sin(3\theta+\varphi)\right]~~~
\end{eqnarray}
with
$E_{Z}=\mu_{B}g_{1}B/2$ and $g_{\pm}=1\pm{g_{2}/g_{1}}$.

\begin{figure}
\includegraphics [width=1.0\columnwidth, keepaspectratio] {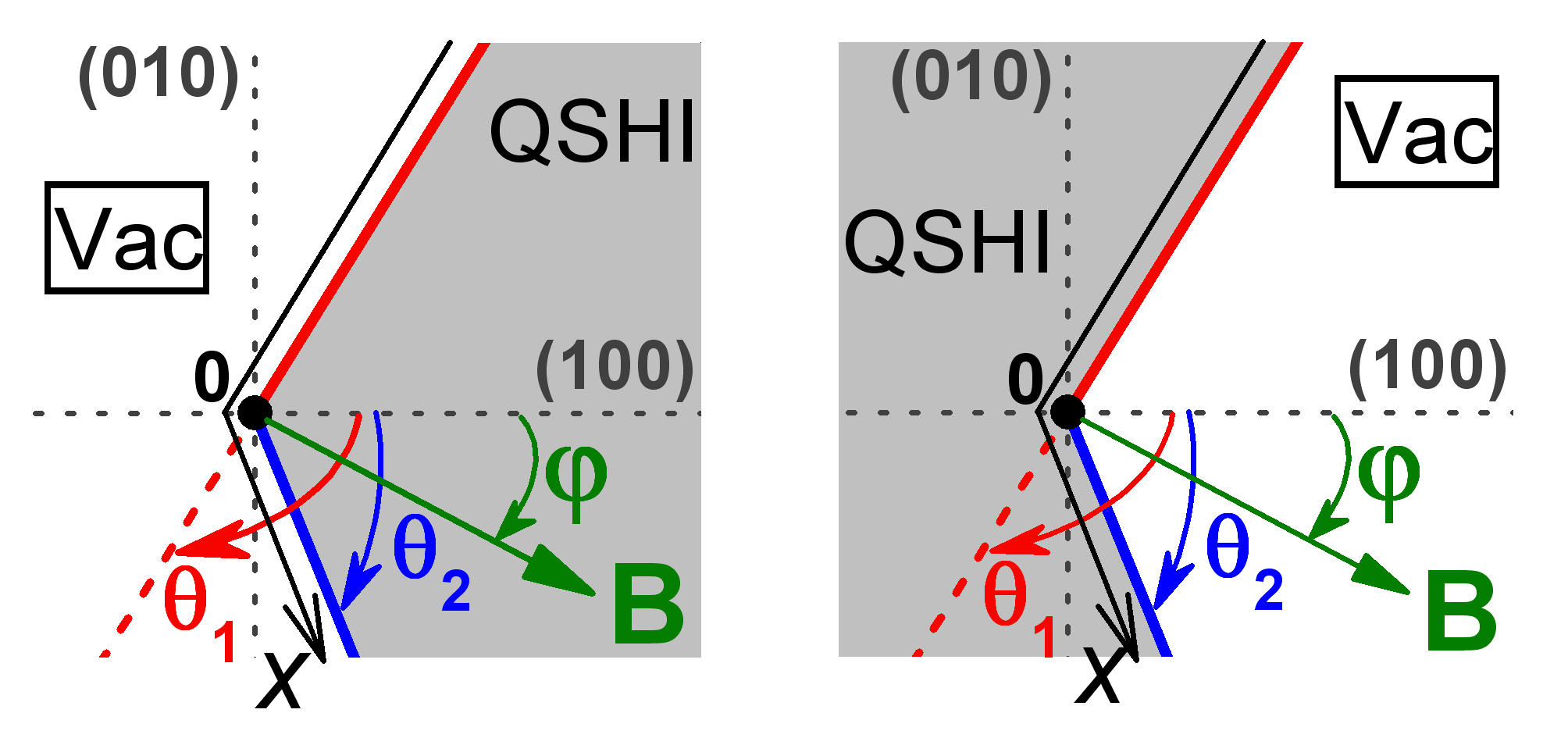} 
\caption{\label{Fig:2} Schematic of orientation of two meeting edges (defined by $\theta_1$ and $\theta_2$) and magnetic field (defined by $\varphi$) with respect to main crystallographic axes in the QW plane. Here, the ``curved'' $x$ axis is oriented along the edges so that $x=0$ represents to the meeting corner, while $x<0$ and $x>0$ correspond to $\theta_1$ and $\theta_2$, respectively. Note that the latter is fulfilled only if $\theta_1-\theta_2\neq\pm\pi$. The QSHI and external vacuum are shown in grey and white, respectively. The positive values of the angles correspond to a clockwise rotation around $z$ axis. The right and left panels represent the same case considered in the text.}
\end{figure}

The obtained edge Hamiltonian provides a convenient starting point for analyzing the junction of two edges with different crystallographic
orientations. Such a junction forms a corner of the sample, where the orientation-dependent mass terms may change discontinuously. The geometry considered in the following analysis is illustrated in Fig.~\ref{Fig:2}.

\section{\label{Sec:Corner} Corner states from a two-mass Dirac theory}
Let us now redefine the coordinate $x$ along the ``curved'' edge so that $x=0$ corresponds to the meeting corner. We model the corner as a junction of two semi-infinite edge segments ($x<0$ and $x>0$) characterized by different constant mass vectors $(M_x,M_y)$ far from the corner. The latter means that $M_x$ and $M_y$ in Eq.~(\ref{eq:8}) are now the functions of $x$, and $\hat{k}_x=-i\partial/\partial{x}$. Note that the edge Hamiltonian is defined in disjoint regions out of $x=0$. As a boundary condition at $x=0$, we further assume the continuity of the edge wave-function. In view of the above, the Schr\"{o}dinger equation for the corner states takes the form
\begin{multline}
\label{eq:Dirac1}
\left({\hbar}v_{F}\hat{k}_x{s}_z+M_x(x){s}_x+M_y(x){s}_y\right)\Psi_{\mathrm{0D}}(x)=\\
=\left(E-\varepsilon_0\right)\Psi_{\mathrm{0D}}(x).
\end{multline}

Let us act by the matrix operator from the left-hand side of Eq.~(\ref{eq:Dirac1}) on both sides of this equation. This leads to
\begin{multline}
\label{eq:Dirac2}
\bigg\{\left({\hbar}^{2}v_{F}^2\hat{k}_x^2+M_x^2+M_y^2-\left(E-\varepsilon_0\right)^2\right)+\\
+{\hbar}v_{F}\begin{pmatrix}
0 & -i{M_x}'-M_y' \\
i{M_x}'-M_y' & 0
\end{pmatrix}
\bigg\}\Psi_{\mathrm{0D}}(x)=0,
\end{multline}
where the prime denotes the derivative with respect to $x$.

An exact solution of Eq.~(\ref{eq:Dirac2}) can be found in the case of a \emph{sharp} corner at $x=0$, when
\begin{equation}
\label{eq:Dirac3}
M_x(x)={\alpha_x}\Theta(x)+m_x,~~~~~~~~M_y(x)={\alpha_y}\Theta(x)+m_y,
\end{equation}
where $\Theta(x)$ is a step-function defined as $\Theta(x)=-1$ if $x<0$ and $\Theta(x)=1$ for $x\geq0$, while the constants ${\alpha_x}$, ${\alpha_y}$, ${m_x}$ and ${m_y}$ are written as
\begin{eqnarray}
\label{eq:Dirac4}
\alpha_{x,y}=\dfrac{M_{x,y}(+\infty)-M_{x,y}(-\infty)}{2},\nonumber\\
m_{x,y}=\dfrac{M_{x,y}(+\infty)+M_{x,y}(-\infty)}{2}.
\end{eqnarray}
Hence, solutions of Eq.~(\ref{eq:Dirac2}) is constructed as follows:
\begin{equation}
\label{eq:Dirac6}
\Psi_{\mathrm{0D}}(x)={\chi}\psi(x),
\end{equation}
where ${\chi}$ is the spin part of the wave function satisfying the equation
\begin{equation*}
\begin{pmatrix}
0 & -\alpha_y-i{\alpha_x} \\
-\alpha_y+i{\alpha_x} & 0
\end{pmatrix}{\chi}=\nu{\chi},
\end{equation*}
with the eigenvalues $\nu=\pm\sqrt{\alpha_x^2+\alpha_y^2}$.

By introducing a new variable $\tilde{x}=x/({\hbar}v_{F})$, the equation for the coordinate part $\psi(x)$ can be written as
\begin{equation}
\label{eq:Dirac8}
\Bigg\{\hat{\tilde{k}}_x^2
+{W}(\tilde{x})^2+\sigma{W}(\tilde{x})'\Bigg\}\psi(\tilde{x})=\varepsilon{\psi}(\tilde{x}),
\end{equation}
where $\sigma=\pm{1}$ (the sign of $\sigma$ coincides with those for $\nu$), and $\varepsilon$ and ${W}(\tilde{x})$ are defined as
\begin{eqnarray}
\label{eq:Dirac9}
\varepsilon=\left(E-\varepsilon_0\right)^2-\dfrac{\left(m_x\alpha_y-m_y\alpha_x\right)^2}{\alpha_x^2+\alpha_y^2},\nonumber\\
{W}(\tilde{x})=\dfrac{\alpha_xM_x(\tilde{x})+\alpha_yM_y(\tilde{x})}{\sqrt{\alpha_x^2+\alpha_y^2}}.~~~~
\end{eqnarray}

Importantly, Eq.~(\ref{eq:Dirac8}) mimics the conventional Schr\"{o}dinger equation with an electrostatic potential being a linear combination of the square and the derivative of the same function ${W}(\tilde{x})$. It possesses a special symmetry and represents the formulation of supersymmetric quantum mechanics~\cite{cst58}, which allows for the factorization:
\begin{equation}
\label{eq:Dirac10}
\left(-i\hat{\tilde{k}}_x-\sigma{W}(\tilde{x})\right)\left(i\hat{\tilde{k}}_x-\sigma{W}(\tilde{x})\right){\psi}(\tilde{x})=\varepsilon{\psi}(\tilde{x}),
\end{equation}
If the signs of the asymptotics ${W}(+\infty)$ and ${W}(-\infty)$ are opposite, Eq.~(\ref{eq:Dirac10}) admits a localized solution ${\psi}(\tilde{x})$ with $\varepsilon=0$, which converts the second brackets into zero:
\begin{equation}
\label{eq:Dirac12}
\left(\dfrac{d}{d\tilde{x}}-\sigma{W}(\tilde{x})\right){\psi}(\tilde{x})=0.
\end{equation}
Thus, under this condition, Eq.~(\ref{eq:Dirac1}) always has a localized solution at $x=0$ with the wave-function
\begin{equation}
\label{eq:9}
\Psi_{\mathrm{0D}}(x)=\mathcal{C}
\begin{pmatrix}
\alpha_{y}+i\alpha_{x} \\
-\sigma\sqrt{\alpha_x^2+\alpha_y^2}
\end{pmatrix}
e^{\displaystyle{\frac{\sigma}{{\hbar}v_{F}}\int\limits_0^{x}
W(z)dz}},
\end{equation}
where $\mathcal{C}$ is the normalization constant. The value of $\sigma$ in Eq.~(\ref{eq:9}) should be chosen in accordance with normalized condition of $\Psi_{\mathrm{0D}}(x)$: if $W(+\infty)>0$, $\sigma=-1$, while for $W(+\infty)<0$, $\sigma=1$.

The energy of this 0D corner state always lying in the gap for the edge states is written as
\begin{equation}
\label{eq:11}
E_{\mathrm{0D}}=\mathbb{C}-\eta\mathbb{M}+
\dfrac{{\sigma}\left(\alpha_{x}m_y-\alpha_{y}m_x\right)}{\sqrt{\alpha_x^2+\alpha_y^2}},
\end{equation}
where we took into account that $\varepsilon_0=\mathbb{C}-\eta\mathbb{M}$. As clear from Eq.~(\ref{eq:9}), in the case of two mass parameters, the corner state existence is related to the sign changing of $W(x)$ and not to the band-gap closing upon passing through $x=0$. Unlike the standard Jackiw--Rebbi problem~\cite{cst42} with a single mass term, the presence of two independent mass terms generically shifts the bound-state energy away from a non-zero energy (``zero energy'' in our notations corresponds to $E_{\mathrm{0D}}=\mathbb{C}$). We note that a localized state with the energy $E_{0D}$ in Eq.~(\ref{eq:11}) is the only one state arising at the sharp meeting corner~\cite{SM}.

For further analysis, instead of Eq.~(\ref{eq:8}), it is convenient to apply the following parametrization for $M_x$ and $M_y$:
\begin{equation}
\label{eq:12}
M_x=M\cos{\beta},~~~~~~~~~M_y=M\sin{\beta},
\end{equation}
where $M>0$ and $\beta\in(-\pi,\pi]$ are both functions of $\theta$ and $\varphi$ being defined as $\tan{\beta}=M_{y}/M_{x}$ and
\begin{equation}
\label{eq:13}
M=\sqrt{M_{x}^2+M_{y}^2}=E_{Z}\sqrt{1-g_{+}g_{-}\sin^{2}\left(\theta+\varphi\right)}.
\end{equation}

Expressing the result in terms of the mass magnitude $M$ and the polar angle $\beta$ of the mass vector, we obtain compact formulas for the bound-state energy and the existence condition:
\begin{multline*}
E_{\mathrm{0D}}=\mathbb{C}-\eta\mathbb{M}-\\
-\dfrac{{\sigma}\sqrt{M_{1}M_{2}}\sin({\beta_1-\beta_2})}{2\sqrt{\sin^2\left(\dfrac{\beta_1-\beta_2}{2}\right)+\dfrac{\left(M_1-M_2\right)^2}{4M_{1}M_{2}}}},
\end{multline*}
\begin{multline}
\label{eq:13b}
W(-\infty)W(+\infty)=-M_{1}M_{2}\times\\
\times
\dfrac{\sin^4\left(\dfrac{\beta_1-\beta_2}{2}\right)-\dfrac{\left(M_1-M_2\right)^2}{4M_{1}M_{2}}\cos\left(\beta_1-\beta_2\right)}{\sin^2\left(\dfrac{\beta_1-\beta_2}{2}\right)+\dfrac{\left(M_1-M_2\right)^2}{4M_{1}M_{2}}},
\end{multline}
where indices $1$ and $2$ correspond to the parameters at $x\rightarrow-\infty$ and $x\rightarrow+\infty$, respectively. It can be shown analytically that for $\left(\beta_1-\beta_2\right)\in(-\pi,\pi]$, the existence condition $W(-\infty)W(+\infty)<0$ is satisfied if
\begin{equation}
\label{eq:14}
\left|\beta_1-\beta_2\right|>\arccos\left[\dfrac{\min\{M_{1},M_{2}\}}{\sqrt{M_{1}M_{2}}}\right].
\end{equation}
Since $\beta$ does not coincide with $\theta-\varphi$ when $g_h^{||}$ and $\Delta$ are both non-zero (see Eq.~(\ref{eq:8})), the existence condition  in the general case is fulfilled only for specific orientations of magnetic field and the meeting edges. Equation~(\ref{eq:14}) shows that the corner state appears, when the mass vectors on the two edges are sufficiently misaligned; in the isotropic case $M_1=M_2$ this reduces to $|\beta_1-\beta_2|>0$.

Interestingly, if we ``switch off'' the crystal symmetry of the prototype QW by assuming $g_1=g_2$ (see Eq.~(\ref{eq:6})), $M=E_Z$ and $\beta=\theta-\varphi$, the corner state still exist. Indeed, under such condition, $E_{\mathrm{0D}}$ and $W(-\infty)W(+\infty)$ take the simple form
\begin{eqnarray}
\label{eq:15}
E_{\mathrm{0D}}=\mathbb{C}-\eta\mathbb{M}-\sigma\left|E_Z\right|\cos\left(\dfrac{\theta_1-\theta_2}{2}\right),~~~\nonumber\\
W(-\infty)W(+\infty)=-E_Z^2\sin^2\left(\dfrac{\theta_1-\theta_2}{2}\right)<0.
\end{eqnarray}
The latter is fulfilled if $\theta_1\neq\theta_2$, that corresponds to the presence of realistic corner in 2D system (see Fig.~\ref{Fig:2}). In this case, the corner state energy is independent of the magnetic field orientation. Obviously, Eqs.~(\ref{eq:15}) are valid for any QSHIs with isotropic g-factor of the edge states.

\begin{figure}
\includegraphics [width=1.0\columnwidth, keepaspectratio] {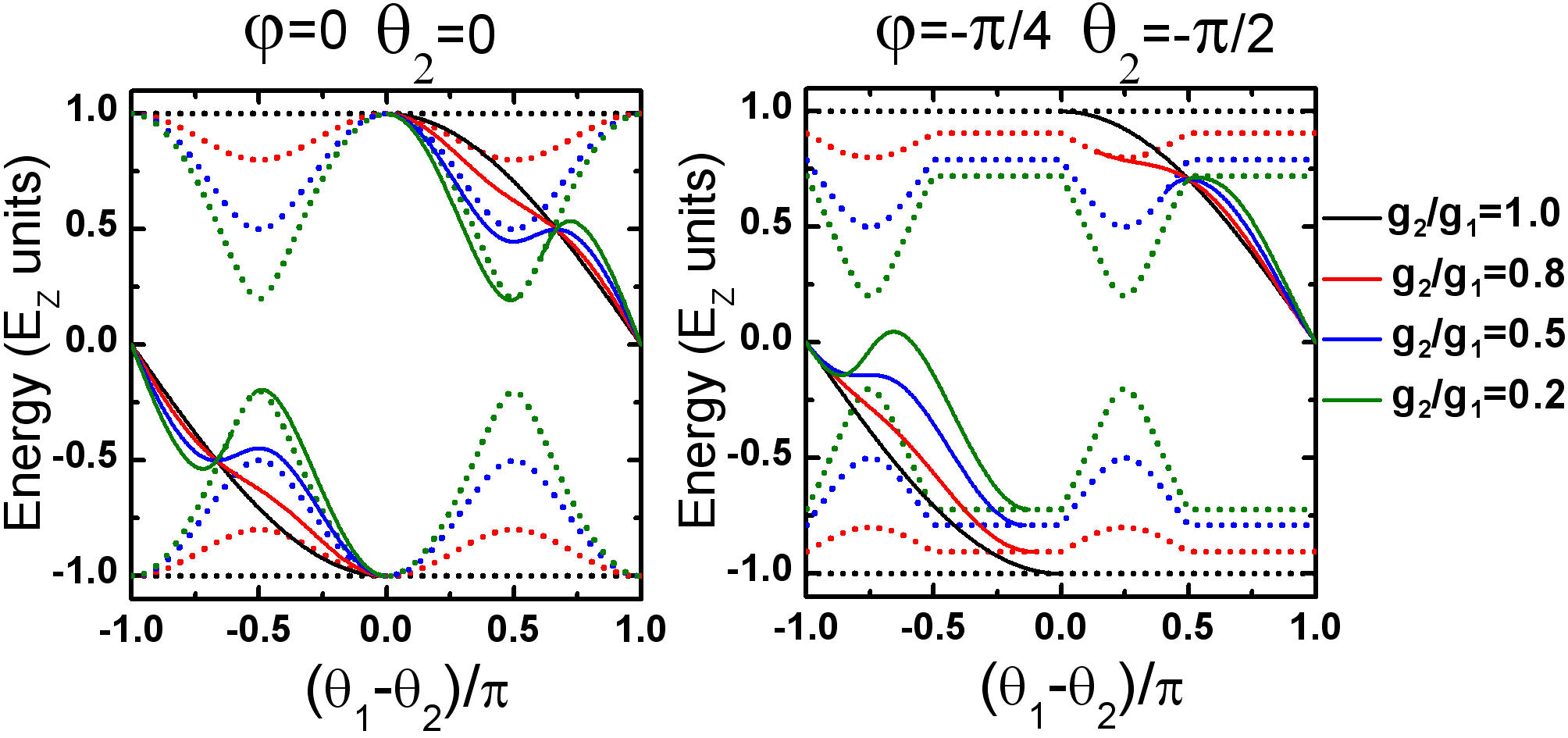} 
\caption{\label{Fig:3} Energy of corner states $E_{\mathrm{0D}}-\varepsilon_0$ (solid curves) for different orientations of the magnetic field and edges relative to the main crystallographic axes, calculated for various ratio of $g_2/g_1$. We remind that $\varepsilon_0=\mathbb{C}-\eta\mathbb{M}$. The dotted curves represented by $\pm\min\{M_{1},M_{2}\}$ correspond to the boundaries of 1D edge band states projected onto the local corner spectrum. The corner state arises as soon as $W(-\infty)W(+\infty)<0$. Note that $\theta_1-\theta_2\neq0$ and $\pm\pi$ for physically reasonable edges forming a common corner.}
\end{figure}

Figure~\ref{Fig:3} shows the evolution of corner state energy as a function of $\theta_1-\theta_2$ at various ratio of $g_2/g_1$, calculated for several orientations of the magnetic field and edges. As seen, the corner state always appears from the gapped 1D edge bands. Each time the bound-state energy approaches the edge continuum, the product $W(-\infty)W(+\infty)$ approaches zero, which signals delocalization of the corner state (cf. Eq.~(\ref{eq:9})). The boundaries of the 1D edge continuum projected onto the local corner spectrum are determined by $\pm\min\{M_{1},M_{2}\}$ shown by dotted curves. Since both $H_{\mathrm{2D}}$ in Eq.~(\ref{eq:1}) and $H_{Z}$ in Eq.~(\ref{eq:3}) do not have electron-hole symmetry, the corner state energy $E_{\mathrm{0D}}$ depends on the orientation of the edges and the magnetic field. Interestingly, at specific values of $g_2/g_1$, $\theta_1$, $\theta_2$ and $\varphi$, the corner state may have a ``zero energy'' (which is $E_{\mathrm{0D}}=\mathbb{C}$ in our notations) coinciding with the corner state energy in the system with particle-hole symmetry. However, this is the result of an arbitrary coincidence for a specific set of parameters , which are in no way related to any particular geometry of the edge and magnetic field orientations (except the case of $g_e^{||}=g_h^{||}$ and $\mathbb{D}=0$~\cite{SM}).

Our general analytical results represented by Eqs.~(\ref{eq:Dirac9})--(\ref{eq:11}) can also be applied for other QSHIs, for which the edge g-factor tensor is different from the one in Eq.~(\ref{eq:5}). For instance, the edge g-factor tensor can be derived from the lowest-order in-$\mathbf{k}$ expansion of tight-binding Hamiltonians widely used to describe QSHIs within the lattice models~\cite{cst47a00,cst47,cst48}.
As can be seen from Supplementary Materials~\cite{SM}, the analytical results obtained on the basis of Eq.~(\ref{eq:11}) with the corresponding edge g-factor tensor perfectly reproduce the results of numerical calculations performed on the square lattice~\cite{cst47a00}.

Let us briefly discuss the effect of a perpendicular magnetic field $B_z$ in zinc-blende semiconductor QWs. The presence of BIA or IIA leads to the band-gap opening of the edge states even for the small values of $B_z$~\cite{cst56}. In this case, the terms describing the band-gap opening are written as
\begin{equation}
\label{eq:16}
H_{\mathrm{edge}\bot}^{(\mathbf{B})}=\dfrac{\mu_{B}B_z}{2}\left(g_{xz}{s}_{x}+g_{yz}{s}_{y}\right),
\end{equation}
where $g_{xz}$ and $g_{yz}$ have the following form:
\begin{equation}
\label{eq:17}
g_{xz}=g_3\sin\theta,~~~~~~g_{yz}=-g_3\cos\theta.
\end{equation}
Here, $g_3\sim\Omega^3\Delta$ is a non-zero constant~\cite{cst56} due to the presence of BIA or IIA in the zinc-blende QWs. Note that the diagonal component of the edge g-factor $g_{zz}$ is independent of $\theta$ and, thus, it can be formally nullified using the substitution $\hat{k}_x\rightarrow\hat{k}_x-g_{zz}\mu_{B}B_z/(2{\hbar}v_{F})$.

As clear, the form of $H_{\mathrm{edge}\bot}^{(\mathbf{B})}$ in Eq.~(\ref{eq:16}) coincides with those for Hamiltonian~(\ref{eq:7}). This means that out-of-plane magnetic field also leads to the corner state arising at the intersection of two meeting edges. Indeed, using the previously obtained results, the existence condition and energy of the corner state are directly written as
\begin{eqnarray}
\label{eq:18}
W(-\infty)W(+\infty)=-\dfrac{\mu_{B}^2g_{3}^2B_{z}^2}{4}\sin^2\left(\dfrac{\theta_1-\theta_2}{2}\right)<0,\nonumber\\
E_{\mathrm{0D}\bot}=\mathbb{C}-\eta\mathbb{M}-\sigma\dfrac{\left|\mu_{B}g_{3}B_{z}\right|}{2}
\cos\left(\dfrac{\theta_1-\theta_2}{2}\right).~~~~
\end{eqnarray}
As clear from Eq.~(\ref{eq:18}), in the presence of BIA or IIA when $g_3\neq{0}$, the corner state induced by out-of-plane magnetic field exists for any crystallographic orientations of the meeting edges.

Interestingly, the parametrization in Eq.~(\ref{eq:12}) shows that the pair $(M_x,M_y)$ can be viewed as a two-component mass vector in the $(s_x,s_y)$ plane, with magnitude $M$ and polar angle $\beta$. For the two edges meeting at the corner, the corresponding asymptotic mass vectors can therefore be written as
\begin{equation}
\label{eq:18a}
\mathbf{M}_{1,2}=M_{1,2}\hat{\mathbf{m}}_{1,2}, \qquad
\hat{\mathbf{m}}_{1,2}=(\cos\beta_{1,2},\sin\beta_{1,2}),
\end{equation}
where $\hat{\mathbf{m}}_{1,2}$ are the unit mass vectors. In this notation, the corner-state energy in Eq.~(\ref{eq:13b}) takes the compact form
\begin{equation}
\label{eq:18b}
E_{\mathrm{0D}}=\varepsilon_0+\sigma\dfrac{\left(\mathbf{M}_1\times\mathbf{M}_2\right)_z}
{\left|\mathbf{M}_1-\mathbf{M}_2\right|},
\end{equation}
where $\left(\mathbf{M}_1\times\mathbf{M}_2\right)_z=M_1M_2\sin(\beta_2-\beta_1)$.

Similarly, the existence condition for the corner state can be written in terms of the relative orientation of the unit mass vectors as
\begin{equation}
\label{eq:18c}
\hat{\mathbf{m}}_1\cdot\hat{\mathbf{m}}_2=\cos(\beta_1-\beta_2)
<\sqrt{\dfrac{\min\{M_1,M_2\}}{\max\{M_1,M_2\}}}.
\end{equation}
The latter is just a geometric rewriting of Eq.~(\ref{eq:14}). Indeed, for $\Delta\beta=\beta_1-\beta_2\in(-\pi,\pi]$, the condition
$|\Delta\beta|>\arccos r$ is equivalent to $\cos(\Delta\beta)<r$, while
\begin{equation*}
r=\dfrac{\min\{M_1,M_2\}}{\sqrt{M_1M_2}}=
\sqrt{\dfrac{\min\{M_1,M_2\}}{\max\{M_1,M_2\}}}.
\end{equation*}

Equation~(\ref{eq:18c}) shows that the corner state appears when the mass vector rotates sufficiently strongly between the two edges. The required rotation angle is controlled by the ratio of the edge gaps: the stronger the mismatch between $M_1$ and $M_2$, the larger the misalignment between $\hat{\mathbf{m}}_1$ and $\hat{\mathbf{m}}_2$ must be. This geometric picture also makes it transparent how the conventional Jackiw--Rebbi result~\cite{cst42} is recovered. In the standard one-mass Dirac problem, the mass changes sign across the domain wall, which in the present notation corresponds to opposite directions of the mass vectors, $\hat{\mathbf{m}}_2=-\hat{\mathbf{m}}_1$, i.e. $\beta_2-\beta_1=\pi$. In this limit, Eq.~(\ref{eq:18c}) is automatically satisfied, while Eq.~(\ref{eq:18b}) yields $E_{\mathrm{0D}}=\varepsilon_0$. Thus, the conventional Jackiw--Rebbi mode is reproduced as a special case in which the mass vector undergoes a $\pi$ rotation, resulting in
\begin{equation}
\label{eq:18d}
E_{\mathrm{0D}}=\mathbb{C}-\eta\mathbb{M}.
\end{equation}
Hence, for $\eta=0$ (i.e. $\mathbb{D}=0$), the corner state becomes ``zero-energy'', coinciding with the middle of the gap. As will be discussed below,
the situation $\hat{\mathbf{m}}_2=-\hat{\mathbf{m}}_1$ is realized only for corners related to the mirror-symmetric directions.

\section{Corner states and mirror-graded winding numbers}
The geometric interpretation derived at the end of the previous section shows that the corner-bound states in zinc-blende QSHIs are governed by the relative orientation of the edge mass vectors and therefore may arise for a broad range of edge and magnetic-field orientations. It is then natural to ask whether these states can nevertheless be related to bulk topological invariants in those special symmetry settings where such invariants can be defined. In 2D systems with additional mirror symmetry, the corresponding bulk characterization is commonly formulated in terms of mirror-graded winding numbers~\cite{bookSM1q,cst60}. These quantities are commonly associated with higher-order topology in mirror-symmetric settings, where the bulk Hamiltonian admits a decomposition into independent mirror sectors. The key question for the present problem is whether the magnetic-field-induced corner states obtained in the previous section are indeed controlled by such bulk invariants. To identify the special symmetry settings in which mirror-graded winding numbers can be defined, we consider the bulk Hamiltonian along a mirror-invariant line in momentum space. Without loss of generality, we choose the in-plane magnetic field to be oriented along the $y$ axis. In this case, if the crystallographic orientation is such that the bulk Hamiltonian preserves the mirror-reflection symmetry $\mathcal{M}_y$, the line $k_y=0$ remains invariant under this symmetry. We therefore analyze the Hamiltonian ${H_{\mathrm{2D}}}(k_x,0)+{H_{\mathrm{Z}}}$ in the mirror-adapted coordinate system, where the in-plane magnetic field is represented by its $y$ component $B_y$.

In contrast to the zinc-blende case, the lattice models considered in Refs.~\cite{cst47a00,cst47,cst48} do not rely on the crystal-anisotropic edge $g$-factor structure derived from Eqs.~(\ref{eq:5}) and (\ref{eq:6}). Here, by contrast, the low-energy Hamiltonian of zinc-blende semiconductor QWs is not rotationally invariant because of the BIA/IIA term $\Delta$ in Eq.~(\ref{eq:1}) and the anisotropic Zeeman coupling in Eq.~(\ref{eq:3}). Nevertheless, for certain crystallographic directions the bulk Hamiltonian preserves mirror-reflection symmetry, which makes it possible to define mirror-resolved 1D problems and to evaluate the corresponding mirror-graded winding numbers. To make this symmetry reduction explicit, it is convenient to apply the unitary transformation
\begin{equation}
\label{eq:MGW1}
\mathcal{U}=\mathcal{U}'\mathcal{S},
\end{equation}
where
\begin{equation}
\label{eq:MGW2}
\mathcal{U}'=\dfrac{1}{2}\begin{pmatrix}
1 & 1 & -i & -i \\
-1 & 1 & -i & i \\
-i & -i & 1 & 1 \\
-i & i & -1 & 1
\end{pmatrix},
\end{equation}
and
\begin{equation}
\label{eq:MGW3}
\mathcal{S}=\dfrac{1}{\sqrt{2}}\begin{pmatrix}
1 & -i & 0 & 0 \\
-i & 1 & 0 & 0 \\
0 & 0 & 1 & -i \\
0 & 0 & -i & 1
\end{pmatrix}.
\end{equation}
This transformation diagonalizes the mirror-reflection operator
$\mathcal{M}_y$ and brings the Hamiltonian to the form
\begin{widetext}
\begin{multline}
\label{eq:MGW4}
{\mathcal{U}}^{\dag}\left\{{H_{\mathrm{2D}}}(k_x,0)+{H_{\mathrm{Z}}}\right\}{\mathcal{U}}=\mathbb{C}+
\begin{pmatrix}
h_0^{(-)}+h_x^{(-)}\sigma_x+h_y^{(-)}\sigma_y & 0 \\
0 & h_0^{(+)}+h_x^{(+)}\sigma_x+h_y^{(+)}\sigma_y
\end{pmatrix}
+\Delta\cos{2\theta}\begin{pmatrix}
0 & 0 & i & 0 \\
0 & 0 & 0 & -i \\
-i & 0 & 0 & 0 \\
0 & i & 0 & 0
\end{pmatrix}+\\
+\dfrac{\mu_{B}{B_y}}{4}g_e^{||}\cos{(\theta-\varphi)}\begin{pmatrix}
0 & 0 & 1 & 1 \\
0 & 0 & 1 & 1 \\
1 & 1 & 0 & 0 \\
1 & 1 & 0 & 0
\end{pmatrix}
+\dfrac{\mu_{B}{B_y}}{4}g_h^{||}\cos{(5\theta-\varphi)}\begin{pmatrix}
0 & 0 & 1 & -1 \\
0 & 0 & -1 & 1 \\
1 & -1 & 0 & 0 \\
-1 & 1 & 0 & 0
\end{pmatrix},
\end{multline}
where
\begin{eqnarray}
\label{eq:MGW5}
h_0^{(\pm)}&=&\mp\dfrac{\mu_{B}{B_y}}{4}\left[g_e^{||}\sin{(\theta-\varphi)}+g_h^{||}\sin{(5\theta-\varphi)}\right]-\mathbb{D}k_x^2,\nonumber\\
h_x^{(\pm)}&=&\mathbb{M}\mp\dfrac{\mu_{B}{B_y}}{4}\left[g_e^{||}\sin{(\theta-\varphi)}-g_h^{||}\sin{(5\theta-\varphi)}\right]-\mathbb{B}k_x^2,\nonumber\\
h_y^{(\pm)}&=&\Delta\sin{2\theta}\pm \mathbb{A}k_x.
\end{eqnarray}
\end{widetext}
Equation~(\ref{eq:MGW4}) reduces to a block-diagonal form, with the upper and lower blocks corresponding to the $-i$ and $i$ eigenvalues of $\mathcal{M}_y$, respectively, if $\theta=\theta^{*}$ and $\varphi=\varphi^{*}$, where $\theta^{*}=\pm{\pi}/4$ (or $\pm{3\pi}/4$) and $\varphi^{*}-\theta^{*}=\pm{\pi}/2$ (or $\pm{3\pi}/2$) for a non-zero $\Delta$. As seen from Fig.~\ref{Fig:1}, at these $\theta^{*}$ values the line $k_y=0$ corresponds to the intersections of the $(110)$ and $(\bar{1}10)$ mirror-reflection planes with the QW plane. These crystallographic planes are mirror-reflection planes for (001)-grown zinc-blende QWs with $D_{2d}$ symmetry. Note that the magnetic field in this case is oriented perpendicular to the corresponding mirror-symmetric line.

Having identified the special mirror-symmetric geometry in the bulk problem, it is instructive to compare it with the geometric condition $\hat{\mathbf{m}}_2=-\hat{\mathbf{m}}_1$, resulting in $E_{\mathrm{0D}}=\mathbb{C}-\eta\mathbb{M}$, discussed in the previous section. Substituting $\theta=\theta^{*}$ and $\varphi=\varphi^{*}$ with $\varphi^{*}-\theta^{*}=\pm\pi/2$ in Eq.~(\ref{eq:8}), one finds
\begin{equation}
\label{eq:MGW6a}
M_x(\theta^{*},\varphi^{*})=0,\qquad
M_y(\theta^{*},\varphi^{*})=\pm E_Z.
\end{equation}
Thus, the edge mass vector is forced to lie along the $M_y$ axis in the mass plane. Comparing with the geometric discussion in the previous section, this is precisely the special setting in which the mass vectors
on two meeting edges may become antiparallel, $\hat{\mathbf m}_2=-\hat{\mathbf m}_1$, leading for $\eta=0$ to the ``zero-energy'' corner state. In this sense, the midgap corner-state configuration is tied to the corners, whose edges are related to the mirror-symmetric directions.

Returning now to the evaluation of the mirror-graded winding numbers, we note that in the mirror-symmetric configuration discussed above each block in Eq.~(\ref{eq:MGW4}) is described by an effective 1D Hamiltonian
\begin{equation}
\label{eq:MGW7}
h^{(s)}(k_x)=h_0^{(s)}+h_x^{(s)}\sigma_x+h_y^{(s)}\sigma_y,\qquad s=\pm1,
\end{equation}
for which the winding number can be evaluated as
\begin{equation}
\label{eq:MGW8}
\nu^{(s)}=\dfrac{1}{2\pi}\int\limits_{-\infty}^{+\infty}d{k_x}
\dfrac{h_x^{(s)}\partial_{k_x}h_y^{(s)}-h_y^{(s)}\partial_{k_x}h_x^{(s)}}
{\left[h_x^{(s)}\right]^2+\left[h_y^{(s)}\right]^2}.
\end{equation}

Substituting Eq.~(\ref{eq:MGW5}) into Eq.~(\ref{eq:MGW8}) yields
\begin{widetext}
\begin{equation}
\label{eq:MGW9}
\nu^{(s)}=-\dfrac{1}{2\pi}\Bigg[
\arctan\left\{\dfrac{\mathbb{B}}{s\mathbb{A}}k_x+\dfrac{\mathbb{B}\Delta\sin2\theta^{*}}{\mathbb{A}^2}\right\}
+\arctan\left\{
\dfrac{s\mathbb{A}}{\mathbb{M}^{(s)}}\dfrac{\mathbb{B}^2}{\mathbb{A}^2-\dfrac{\mathbb{B}}{\mathbb{M}^{(s)}}\Delta^2\sin^{2}2\theta^{*}}
k_x^3+\mathcal{F}^{(s)}(k_x)\right\}
\Bigg]\Bigg|_{-\infty}^{+\infty}.
\end{equation}
where
\begin{equation*}
\mathcal{F}^{(s)}(k_x)=\dfrac{\mathbb{B}^2\Delta\sin2\theta^{*}}{\mathbb{A}^2\mathbb{M}^{(s)}-\mathbb{B}\Delta^2\sin^{2}{2\theta^{*}}}k_x^2
+s\mathbb{A}\dfrac{\mathbb{A}^2-\mathbb{B}\mathbb{M}^{(s)}}{\mathbb{A}^2\mathbb{M}^{(s)}-\mathbb{B}\Delta^2\sin^{2}{2\theta^{*}}}k_x -\Delta\sin2\theta^{*}\dfrac{\mathbb{A}^2+\mathbb{B}\mathbb{M}^{(s)}}{\mathbb{A}^2\mathbb{M}^{(s)}-\mathbb{B}\Delta^2\sin^{2}{2\theta^{*}}},
\end{equation*}
\begin{equation}
\label{eq:MGW10}
\mathbb{M}^{(s)}=\mathbb{M}-s\dfrac{\mu_{B}{B_y}}{4}\left[g_e^{||}\sin{(\theta^{*}-\varphi^{*})}-g_h^{||}\sin{(5\theta^{*}-\varphi^{*})}\right].
\end{equation}
\end{widetext}

Assuming $\mathbb{B}<0$ and $\mathbb{M}<0$, which guarantees the QSHI state in the absence of magnetic field in the prototype QW~\cite{TAI3}, and
restricting ourselves to magnetic fields small enough such that $\mathrm{sgn}\,\mathbb{M}^{(s)}=\mathrm{sgn}\,\mathbb{M}$, one obtains
\begin{equation}
\label{eq:MGW11}
\nu^{(s)}=\mathrm{sgn}(s\mathbb{A}).
\end{equation}
Thus, for these special mirror-symmetric settings the mirror-resolved 1D sectors indeed carry quantized winding numbers. This bulk-topological result, however, should be interpreted with care. The quantized values in Eq.~(\ref{eq:MGW11}) are obtained only for the special orientations $\theta=\theta^{*}$ and $\varphi=\varphi^{*}$ for which the bulk Hamiltonian can be reduced to the mirror sectors. By contrast, the corner-state analysis of the previous section shows that localized corner states exist in a much broader range of edge and magnetic-field orientations. In particular, the existence condition in Eq.~(\ref{eq:14}) is governed by the relative rotation of the edge mass vectors and remains satisfied for many corners that do not preserve any mirror symmetry of the bulk states.

Therefore, the corner states obtained from the two-mass Dirac theory are not generally encoded in the mirror-graded winding numbers of the bulk Hamiltonian. The latter characterize only special symmetry-restricted configurations, whereas the corner states themselves are controlled by the edge-mass geometry and persist well beyond those configurations. In other words, mirror-graded winding numbers, whenever they can be defined, capture only a restricted subset of the corner-state configurations and do not provide a general criterion for their existence. This conclusion is fully consistent with the geometric interpretation derived at the end of the previous section: crystal symmetry constrains the possible orientations of the edge mass vectors and, in special cases, allows one to define mirror-resolved winding numbers, yet the corner state itself is generated by the rotation of the magnetic-field-induced edge masses rather than by the value of the bulk invariant. 

Interestingly, in the special symmetry-restricted limit considered in Refs.~\cite{cst47a00,cst47,cst48}, the corner states can be consistently interpreted as higher-order topological modes associated with mirror-graded winding numbers. However, this interpretation relies on an additional spectral symmetry that pins the states to the middle of the gap and does not extend to the generic case with electron-hole asymmetry. As shown in the Supplementary Materials~\cite{SM}, after including electron-hole asymmetry these states generally move away from this special energy while remaining within the same edge-mass description. The same message applies to zinc-blende QSHIs: even in the special limit of $\mathbb{D}=0$ (and hence $\eta=0$), ``zero-energy'' corner states occur only when the mass vectors on the two meeting edges become antiparallel, which in the present system happens only for corners related to the relevant mirror planes. For generic corners, the corner-state energy is shifted away from the middle of the gap. Consequently, magnetic-field-induced corner states are not restricted to the higher-order-topological interpretation tied to the zero-energy limit and should not be identified with corner modes protected by mirror-graded winding numbers.

\section{Spectral robustness of the corner states}
The analysis above shows that magnetic-field-induced corner states in zinc-blende QSHIs are generally not protected by a stable bulk topological invariant. This, however, does not imply that they are spectrally fragile in any practical sense. Rather, the present problem naturally leads to a distinction between \emph{topological protection} and \emph{spectral robustness}. By topological protection we mean that the existence of a boundary state is enforced by a bulk invariant together with the protecting symmetries, so that the state cannot be removed without closing the relevant bulk or boundary gap or breaking the protecting symmetry. By spectral robustness we mean a weaker, but physically important, property: under weak disorder, interactions, or other symmetry-allowed perturbations, the state remains a spectrally identifiable in-gap excitation, although its energy, localization length, and spectral weight may be renormalized. For the corner states considered here, the second notion is the more relevant one. 

A natural language for addressing this question is based on the quasiparticle picture~\cite{TAI3,TAI4}, which describes the low-energy physics in terms of a quasiparticle Hamiltonian $\mathcal{H}_{qp}(\mathbf{k},E)$, introduced through the single-particle Green function $\hat{G}(\mathbf{k},E)$:
\begin{equation}
\label{eq:Spectr1}
\mathcal{H}_{qp}(\mathbf{k},E)\equiv E-\hat{G}^{-1}(\mathbf{k},E)
=H_{\mathrm{2D}}(\mathbf{k})+\hat{\Sigma}(\mathbf{k},E),
\end{equation}
where the self-energy $\hat{\Sigma}(\mathbf{k},E)$ encodes effects of perturbations. Physically, the real part of $\hat{\Sigma}(\mathbf{k},E)$ characterizes renormalization of the band structure, while the imaginary part describes the quasiparticle decay resulting in the finite quasiparticle lifetime. Importantly, the non-Hermitian quasiparticle Hamiltonian $\mathcal{H}_{qp}(\mathbf{k},E)$ can be used not only for topological characterization of the system~\cite{TAI3,TAI4}, but also as a starting point for constructing the corresponding effective edge theory~\cite{TAI3}. 

Although the quasiparticle approach provides a natural framework for treating perturbations encoded in the single-particle Green function, a systematic analysis of generic perturbation effects on the corner-bound states is beyond the scope of the present work. As an illustrative example, we consider here the \emph{spectral robustness} of the magnetic-field-induced corner states with respect to short-range electrostatic disorder. In QSHI systems, such disorder may lead to the topological Anderson insulator state~\cite{TAI1,TAI2}. For simplicity, we also neglect the joint effect of BIA and IIA by setting $\Delta=0$, which allows us to use the results from Ref.~\cite{TAI3}. In this case, the self-energy matrix $\hat{\Sigma}(E)$ becomes diagonal and independent of momentum $\mathbf{k}$, while the quasiparticle Hamiltonian $\mathcal{H}_{qp}(\mathbf{k},E)$ differs from $H_{\mathrm{2D}}(\mathbf{k})$ only through the substitutions $\mathbb{C}\rightarrow\widetilde{\mathbb{C}}(E)$ and $\mathbb{M}\rightarrow\widetilde{\mathbb{M}}(E)$~\cite{TAI3}. Here, both $\widetilde{\mathbb{C}}(E)$ and $\widetilde{\mathbb{M}}(E)$ are complex quantities incorporating the renormalization induced by the self-energy $\hat{\Sigma}(E)$, whereas the other parameters entering $H_{\mathrm{2D}}(\mathbf{k})$ remain unchanged within this approximation~\cite{TAI3}. 

Adding the Zeeman coupling $H_{\mathrm{Z}}$ from Eq.~(\ref{eq:3}), one may then repeat for $\mathcal{H}_{qp}(\mathbf{k},E)$ the same projection procedure that was performed for $H_{\mathrm{2D}}(\mathbf{k})$ in deriving the effective edge Hamiltonian. As a result, the edge physics is again described by an effective Dirac Hamiltonian of the same general form,
\begin{equation}
\label{eq:Spectr2}
\mathcal{H}_{\mathrm{edge}}^{qp}(\hat{k}_x,E)
=\widetilde{\varepsilon}_0(E)+{\hbar}v_{F}\hat{k}_x{s}_z +\widetilde{M}_x(E)s_x +\widetilde{M}_y(E)s_y,
\end{equation}
where $\widetilde{\varepsilon}_0(E)$, $\widetilde{M}_x(E)$, and $\widetilde{M}_y(E)$ are renormalized quantities determined by the self-energy $\hat{\Sigma}(E)$. Importantly, for the short-range disorder, one has $\mathrm{Im}\,\hat{\Sigma}(E)=0$ in the energy range corresponding to the renormalized bulk band gap~\cite{TAI3}. Therefore, $\widetilde{\varepsilon}_0(E)$, $\widetilde{M}_x(E)$, and $\widetilde{M}_y(E)$ remain real. This, in turn, implies that the eigenvalues $E_{\mathrm{edge}}^{qp}(k_x,E,\lambda)$ of the effective edge Hamiltonian
\begin{multline}
\label{eq:Spectr2b}
E^{qp}_{\mathrm{edge}}(k_x,E,\lambda)=\widetilde{\varepsilon}_0(E)+\\
+\lambda\sqrt{{\hbar}^2v_{F}^2k_x^2+{\widetilde{M}_x}^2(E)+{\widetilde{M}_y}^2(E)},
\qquad \lambda=\pm1,
\end{multline}
are also real. As a result, the spectral function of the edge states in the bulk-gap energy range takes the form of a delta function,
\begin{equation}
\label{eq:Spectr2c}
A^{\mathrm{edge}}(k_x,E)=\sum_{\lambda}
\delta\!\left[E-E^{qp}_{\mathrm{edge}}(k_x,E,\lambda)\right].
\end{equation}

In this case, to determine a state localized at the intersection of two meeting edges, each described by $\mathcal{H}_{\mathrm{edge}}^{qp}(\hat{k}_x,E)$, one can directly apply the results obtained in Sec.~\ref{Sec:Corner}. It then follows that $\mathcal{H}_{\mathrm{edge}}^{qp}(\hat{k}_x,E)$ admits a localized solution with the real eigenvalue $\widetilde{E}^{qp}_{\mathrm{0D}}(E)$ in the form
\begin{equation}
\label{eq:Spectr3}
\widetilde{E}^{qp}_{\mathrm{0D}}(E)=\widetilde{\varepsilon}_0(E)+\sigma\dfrac{\left(\widetilde{\mathbf{M}}_1(E)\times
\widetilde{\mathbf{M}}_2(E)\right)_z}{\left|\widetilde{\mathbf{M}}_1(E)-\widetilde{\mathbf{M}}_2(E)\right|},
\end{equation}
where $\widetilde{\mathbf{M}}_{1,2}(E)$ are the renormalized mass vectors on the two meeting edges. The corresponding existence condition is again given by the same geometric criterion as in Eq.~(\ref{eq:18c}), namely
\begin{equation}
\label{eq:Spectr4}
\cos\left(\widetilde{\beta}_1(E)-\widetilde{\beta}_2(E)\right)
<\sqrt{\dfrac{\min\left\{\widetilde{M}_1(E),\widetilde{M}_2(E)\right\}}
{\max\left\{\widetilde{M}_1(E),\widetilde{M}_2(E)\right\}}}.
\end{equation}
Thus, within the quasiparticle description, the spectral robustness of the magnetic-field-induced corner state is reduced to the stability of the renormalized edge-mass configuration. As long as the edge gaps remain open and the relative orientation of the renormalized mass vectors on the two meeting edges continues to satisfy Eq.~(\ref{eq:Spectr4}), the corner state persists as a well-defined in-gap quasiparticle excitation. In this regime, its
spectral function retains the delta-function form
\begin{equation}
\label{eq:Spectr5}
A^{\mathrm{corner}}(E)=
\delta\!\left[E-\widetilde{E}^{qp}_{\mathrm{0D}}(E)\right].
\end{equation}

For a generic perturbation, different from the short-range electrostatic disorder considered above, the effective edge Hamiltonian may become non-Hermitian. In that case, the spectral function of the quasiparticle edge states acquires a finite linewidth, which in the weak-perturbation regime is proportional to the perturbation strength. Accordingly, the corner excitation, if it survives, is described not by an infinitely sharp bound-state level but by a complex quasiparticle pole $\widetilde{E}^{qp}_{\mathrm{0D}}(E)$. In the present context, spectral robustness therefore means that the corner state remains associated with at least a sharp resonance of the corner spectral function,
\begin{equation}
\label{eq:Spectr6}
A^{\mathrm{corner}}(E)=
-\dfrac{1}{\pi}
\dfrac{\mathrm{Im}\,\widetilde{E}^{qp}_{\mathrm{0D}}(E)}
{\left[E-\mathrm{Re}\,\widetilde{E}^{qp}_{\mathrm{0D}}(E)\right]^2+
\left(\mathrm{Im}\,\widetilde{E}^{qp}_{\mathrm{0D}}(E)\right)^2},
\end{equation}
with the linewidth remaining small compared to the edge gap. In this sense, the corner states studied here provide an explicit example of \emph{spectral robustness without topological protection}. Their stability is not guaranteed by a stable bulk invariant, but neither is it merely an accidental consequence of fine tuning. Rather, it follows from the stability of the quasiparticle edge description itself: as long as the effective edge Hamiltonian retains the same Dirac structure and the renormalized masses remain in the bound-state regime, the corner-state excitation persists as a spectrally isolated in-gap excitation.

\section{Conclusions} 
We developed an analytical theory of magnetic-field-induced corner states in QSHIs beyond the particle-hole-symmetric limit. Starting from a realistic continuum model for zinc-blende semiconductor QW, we derived the effective edge Hamiltonian in the presence of BIA/IIA and Zeeman coupling. We showed that the corner-state problem reduces to a two-mass Dirac junction, i.e., to a generalized Jackiw--Rebbi mechanism at the meeting of two edges with different magnetic-field-induced masses. Our main result is that, in realistic zinc-blende QSHIs, these corner states are controlled by the relative configuration of the edge mass vectors and therefore are most naturally understood as bound states of the effective edge theory. In general, they are not higher-order topological corner modes protected by a stable bulk invariant. Although mirror-graded winding numbers can be defined and quantized for certain crystallographic orientations of the edges and magnetic field, the existence of corner states in zinc-blende QSHIs is not restricted to parameter regimes in which these bulk invariants are well defined. In particular, the corner states persist over a much broader range of edge orientations and magnetic-field directions, including situations in which mirror-graded winding numbers vanish or cannot be defined. Finally, we argue that, even without topological protection enforced by a stable higher-order bulk invariant, magnetic-field-induced corner states may remain spectrally robust under weak perturbations. This occurs when the perturbation effects reduce to parameter renormalization of the effective quasiparticle edge Hamiltonian while its Dirac structure is retained, so that the corner state survives as a spectrally isolated in-gap excitation.


\begin{acknowledgments}
This work was supported by the Occitanie region through the programs ``Terahertz Occitanie Platform'' and ``Quantum Technologies Key Challenge'' (TARFEP project). We also acknowledge financial support from the French Agence Nationale pour la Recherche through ``Cantor'' (ANR-23-CE24-0022) and ``Teaser'' (ANR-24-CE24-4830) projects.
\end{acknowledgments}


%

\end{document}